\documentclass[onecolumn,11pt]{article}
\usepackage[top=1in, bottom=1in, left=1in, right=1in]{geometry}

\usepackage{times}
\usepackage{sidecap}
\usepackage{cite}

\usepackage{dcolumn}% Align table columns on decimal point
\usepackage{bm}% bold math
\usepackage{amsmath}
\usepackage{amssymb}
\usepackage{revsymb}
\usepackage{graphicx}
\usepackage{setspace}
\usepackage{color}
\usepackage{wrapfig}
\usepackage{url}
\usepackage[labelfont=bf, labelsep=space]{caption}
\usepackage{lineno}
\usepackage{geometry}
\usepackage{float}
\usepackage{enumitem}

\usepackage{fancyhdr}
\begin{document}

\newpage
% \centerline{\textbf{\LARGE{Scalable self-cleaning and self-cooling white or colorful paper}}}
\centerline{\textbf{\LARGE{Self-cleaning and self-cooling paper}}}
% \centerline{\textbf{\LARGE{Self-cleaning radiative cooling paper}}}

\centerline{\textbf{Yanpei Tian$^{1,\dag}$, Hong Shao$^{2,\dag}$, Xiaojie Liu$^1$, Fangqi Chen$^1$, Yongsheng Li$^2$, Changyu Tang$^{2,*}$, Yi Zheng$^{1,3,*}$}}

\centerline{$^1$Department of Mechanical and Industrial Engineering, Northeastern University, Boston, MA, USA.}

\centerline{$^2$Chengdu Green Energy and Green Manufacturing Technology R\&D Center,}
\centerline{Chengdu Development Center of Science and Technology, Sichuan, China.}
% \centerline{China Academy of Engineering Physics, Chengdu, Sichuan, China.}
% \centerline{$^3$Department of Physics, Brown University, Providence, RI, USA.}
\centerline{$^3$Department of Electrical and Computer Engineering, Northeastern University, Boston, MA, USA.}
\centerline{$^*$Corresponding author. Email: y.zheng@northeastern.edu (Y.Z.); sugarchangyu@163.com (C.T.)}
\centerline{$^{\dag}$These authors contributed equally to this work.}

\vspace{-10pt}
\section*{\large{Abstract}}
\vspace{-8pt}
The technique of passive daytime radiative cooling (PDRC) is used to cool an object down by simultaneously reflecting sunlight and thermally radiating heat to the cold outer space through the Earth's atmospheric window. However, for practical applications, current PDRC materials are facing unprecedented challenges such as complicated and expensive fabrication approaches and performance degradation arising from surface contamination. Here, we develop a scalable paper-based material with excellent self-cleaning and self-cooling capabilities, through air-spraying ethanolic polytetrafluoroethylene (PTFE) microparticles suspensions embedded within the micropores of the paper. The formed superhydrophobic PTFE coating not only protects the paper from water wetting and dust contamination for real-life applications but also reinforces its solar reflectance by sunlight backscattering. The paper fibers, when enhanced with PTFE particles, efficiently reflect sunlight and strongly radiate heat through the atmospheric window, resulting in a sub-ambient cooling performance of 5$^{\circ}$C and radiative cooling power of 104 W/m$^2$ under direct solar irradiance of 834 W/m$^2$ and 671 W/m$^2$, respectively. The self-cleaning surface of the cooling paper extends its lifespan and keep its good cooling performance for outdoor applications. Additionally, dyed papers are experimentally studied for broad engineering applications. They can absorb appropriate visible wavelengths to display specific colors and effectively reflect near-infrared lights to reduce solar heating, which synchronously achieves effective radiative cooling and aesthetic varieties in a cost-effective, scalable, and energy-efficient way. 

\newpage
% \section*{Introduction}
Compressor-based cooling systems, providing comfortable interior environments for artificial structures (e.g., buildings), account for about 20\% of the total electricity consumption around the world in 2019 \cite{electricity2019}. Moreover, the resultant heating effects \cite{bridgman2013nature} and greenhouse gas emissions towards the environment \cite{greenhousegas} accelerate global warming and climate changes. Therefore, energy-efficient and eco-friendly cooling approaches are highly demanded for energy-saving techniques. The emerging passive daytime radiative cooling (PDRC) technique can achieve sub-ambient cooling effects under direct sunlight without any energy consumption by simultaneously reflecting sunlight (0.3$-$2.5 $\mu$m) and radiating excessive heat as infrared thermal radiation to the cold outer space through the atmospheric transparent window (8$-$13 $\mu$m) \cite{raman2014passive,zhai2017scalable,mandal2018hierarchically,li2019radiative}. A typical PDRC material should have both high solar reflectance ($R_{solar}$) and high infrared thermal emittance ($\epsilon_{IR}$) so that a net radiative heat loss can be achieved even under sunlight.\cite{yang2018dual}. Such an approach is becoming an attractive candidate for improving energy efficiencies for buildings because it eliminates the need for coolant, electricity, and compressor required by traditional mechanical cooling systems. 

PDRC structures with high $R_{solar}$ and high $\epsilon_{IR}$ have been widely investigated in recent decades including photonic structures \cite{rephaeli2013ultrabroadband,raman2014passive,chen2016radiative,wang2020scalable}, metallized polymers \cite{gentle2015subambient,kou2017daytime,zhou2019polydimethylsiloxane}, white paints \cite{chen2020colored,bhatia2018passive,mandal2018hierarchically,atiganyanun2018effective}, and dielectric-polymer hybrid metamaterials \cite{zhai2017scalable,zhao2019subambient,gentle2010radiative,yang2018dual}. Design and fabrication of these sub-ambient cooling structures are either complicated or costly. Even worse, they are susceptible to contamination by floating dust, dirt, and soot when exposed to the outdoor environment \cite{lu2015robust}. As a result, surface contamination deteriorates the solar reflectance and cooling performance of these structures (Supplementary Fig. 1). For practical applications, PDRC structures often need frequent cleaning, which is time- and labor-intensive, to keep their unique optical properties. These issues have hindered PDRC structures for real-life applications. To the best of our knowledge, cooling structures with self-cleaning ability have been rarely reported. Furthermore, colored buildings are always desirable for aesthetic purposes \cite{cai2019temperature,lozano2019optical}, so white or silvery surfaces, based on the wideband reflectance in visible wavelengths, confine their wide applications for infrastructures. 

\begin{figure}[!ht]
\centering
\includegraphics[width=1.0\textwidth]{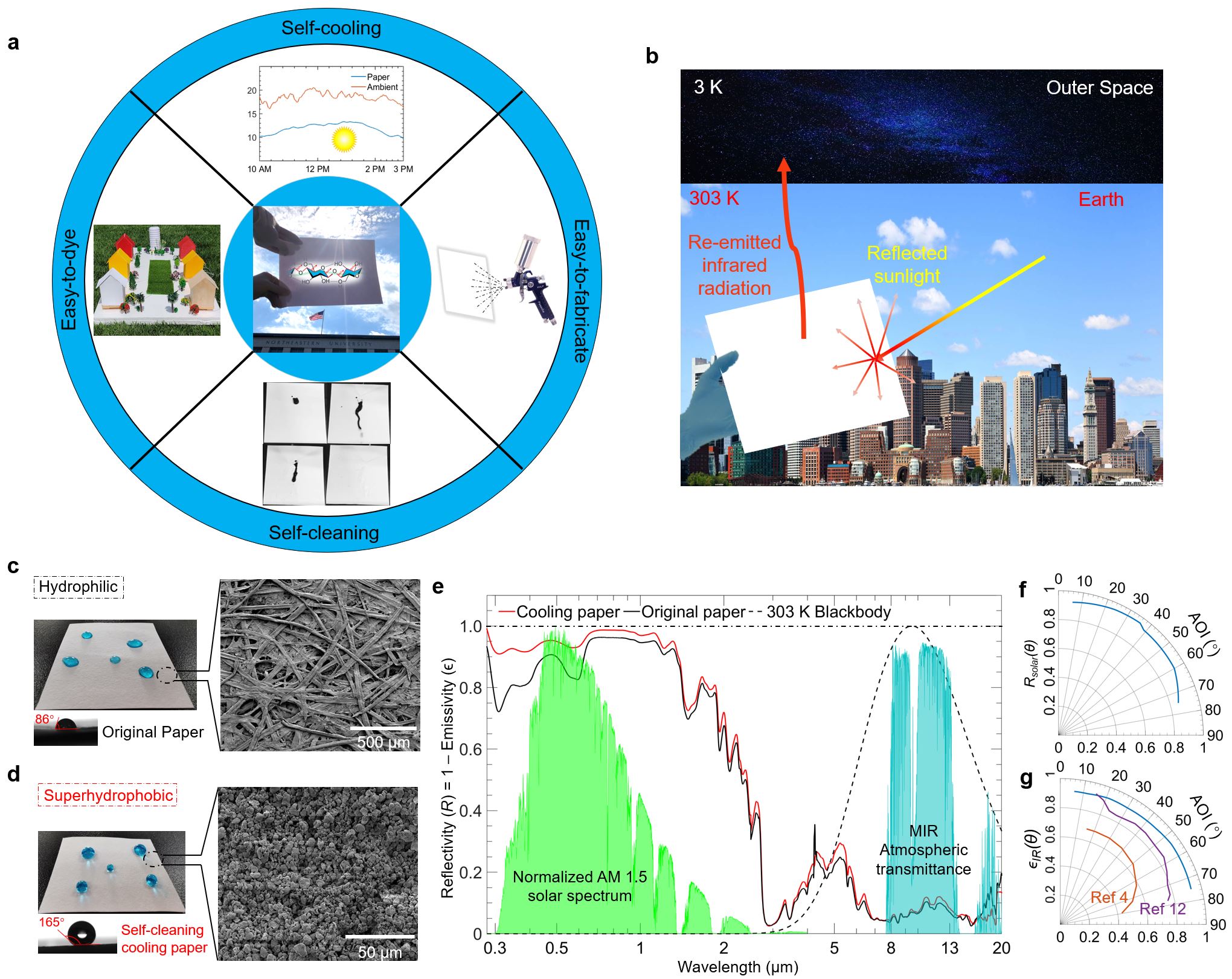}
\caption{\label{fig:mainspectrum} $\mid$ \textbf{Cooling mechanism and surface wetting features of self-cooling and self-cleaning paper.} \textbf{a}, The multifunctional paper with self-cooling and self-cleaning properties, and easy-to-dye and -fabricate features. \textbf{b}, Photograph shows that the cooling paper strongly backscatters the solar irradiance and re-emits thermal radiation, resulting in a net radiative loss to the outer space. \textbf{c}, \textbf{d}, Surface wetting states and micro-structures of original paper and PTFE-coated cooling paper. \textbf{e}, Hemispherical spectral reflectivity ($R$ = 1 $-$ $\epsilon$) of the original and cooling papers displayed with the normalized ASTM G-173 (AM 1.5) solar spectrum, the mid-infrared (MIR) atmospheric transparent window, and the spectral irradiation of a 303 K blackbody. \textbf{f}, \textbf{g}, High solar reflectivity (\textbf{f}, 0.3 $\mu$m $-$ 2.5 $\mu$m) and infrared thermal emissivity (\textbf{g}, 8 $\mu$m $-$ 13 $\mu$m) across angles of incidence (AOI) result in an overall good performance of the cooling paper with close-to-unity, incident angle-independent hemispherical $R_{solar}$ and $\epsilon_{IR}$ that 
are higher than previously reported values (ref. \cite{raman2014passive} and \cite{gentle2015subambient}).
} 
\end{figure}

The cellulose-based paper that is widely used in our daily life has emerged as a low-cost engineering material for various devices (e.g., electronic device and paper-based lab on a chip) due to its earth-abundance, biodegradability, and remarkable optical and mechanical properties \cite{fang2014highly,yamada2017toward}.
The paper consisting of cellulose fibers (20$-$50 $\mu$m in diameter) can strongly scatter sunlight with a high $R_{solar}$ of about 0.89 and meanwhile exhibits an excellent $\epsilon_{IR}$ of 0.92 owing to the molecular vibrations of C-O-C and C-O bonds in cellulose \cite{schwanninger2004effects}. Thus, the paper could be a good candidate for PDRC materials. However, the paper is not water-proof and is easily contaminated by dust, resulting in a significant reduction of its optical structures and cooling performance. Here, we report a simple and scalable method to transform the paper into a self-cooling material with self-cleaning ability \cite{li2007facile}. Polytetrafluoroethylene (PTFE) particles, an ultra-white material with extremely high reflectance (0.3$-$2.5 $\mu$m) \cite{yang2018dual} and extremely low surface energy, are sprayed onto paper to form a white superhydrophobic coating. The water droplets can roll off its surface and take away the dust on the surface presenting good self-cleaning ability. The presence of PTFE coating not only protects the paper from water wetting and dust contamination for outdoor applications but also reinforces its solar reflectance ($\epsilon_{IR}$ $=$ 0.93). Consequently, the PTFE-coated paper exhibits a sub-ambient cooling of $\sim$ 5$^{\circ}$C and a cooling power of $\sim$ 104 W/m$^2$ under solar irradiance of 834 W/m$^2$ and 671 W/m$^2$, respectively. This method can integrate self-cleaning and self-cooling surface to various substrates such as brick, metal, wood, and concrete for versatile applications. The self-cleaning feature ensures that its life cycle is maintenance-free and extends its lifespan. Besides, the recyclability of the cooling paper after one life cycle also contributes to its green and energy-saving functionality (Supplementary Fig. 2a). Furthermore, the cooling paper has diverse dyeing options to attain the necessity of aesthetics along with modest cooling performance. These advantages (Fig. \ref{fig:mainspectrum}a) of the functional cooling paper will shed light on an alternative novel paper-based infrastructure radiative cooling materials and provide a promising pathway of the green-energy applications.

\section*{Surface wetting and optical properties of the functional cooling paper}
The original paper consists of random-stacked cellulose fibers with porous structures (Supplementary Fig. 2b,3). The cooling paper is fabricated by air-spraying homogeneous PTFE microparticles suspension onto the original paper (Fig. \ref{fig:mainspectrum}a and Supplementary Fig. 2-4) followed by solvent evaporation. The PTFE microparticles with an averaged size of 1.4 $\mu$m embed into the micropores (20 $\mu$m) of the paper formed by the stacked cellulose fibers and form a white coating. The random-stacked fibers with the PTFE particles efficiently scatter sunlight from ultraviolet to near-infrared wavelengths and throw heat into the cold outer space (Fig. \ref{fig:mainspectrum}b). Meanwhile, the PTFE particles form a rough surface with micro-nano structures, which transform the paper from hydrophilic (contact angle, $\theta$ = 86$^\circ$) to superhydrophobic state ($\theta$ = 165$^\circ$) (Fig. \ref{fig:mainspectrum}c,d, Supplementary Fig. 5a,b, and Supplementary Video 1). This feature enables the water-proof and self-cleaning surface of the paper for outdoor applications. Interestingly, $R_{solar}$ of the paper is enhanced from 0.89 to 0.93 after coating PTFE, compared with the original paper. This result should be attributed to PTFE microparticles coating that reinforces the sunlight scattering of the paper (Fig. \ref{fig:mainspectrum}e and Supplementary Fig. 5c,d). The high hemispherical solar reflectance (nearly ``white") and infrared emissivity, (nearly ``black") of the cooling paper ensures that the cooling paper radiates a net heat flux to the cold outer space sink as infrared thermal radiation (Fig. \ref{fig:mainspectrum}e). The heat radiated by the cooling paper exceeds the gained heat from the solar absorption, thus leading to sub-ambient cooling effects. In the outdoor environment, sunlight irradiates earth with different incident angles all day and is partly polarized by the dust in the air. These factors can affect the optical response of PDRC materials, which is generally undesirable. Fortunately, both solar reflectance and thermal emittance of our cooling paper are angle-independent even a large angle of incident (AOI, $\theta$ = 75$^\circ$) (Fig. \ref{fig:mainspectrum}f,g, and Supplementary Fig. 6), which is attributed to the porous and rough surface \cite{mandal2018hierarchically}. The randomly-arranged cellulose fibers and PTFE particles give rise to a polarization-independent $R_{solar}$ and $\epsilon_{IR}$ (Supplementary Fig. 7). With these unique structures, the cooling paper shows the diffused ``white" and ``black" features of the cooling paper from 6$^\circ$ to 75$^\circ$ in the solar wavelengths and atmospheric transmittance window, respectively, yielding high $R_{solar}$ and $\epsilon_{IR}$ (greater than 0.9, Fig. \ref{fig:mainspectrum}f,g) when the cooling paper faces to different angles of the sky in the real-life applications.

\section*{Physical mechanism of the cooling paper for radiative cooling}

\begin{figure}[!ht]
\centering
\includegraphics[width=1\textwidth]{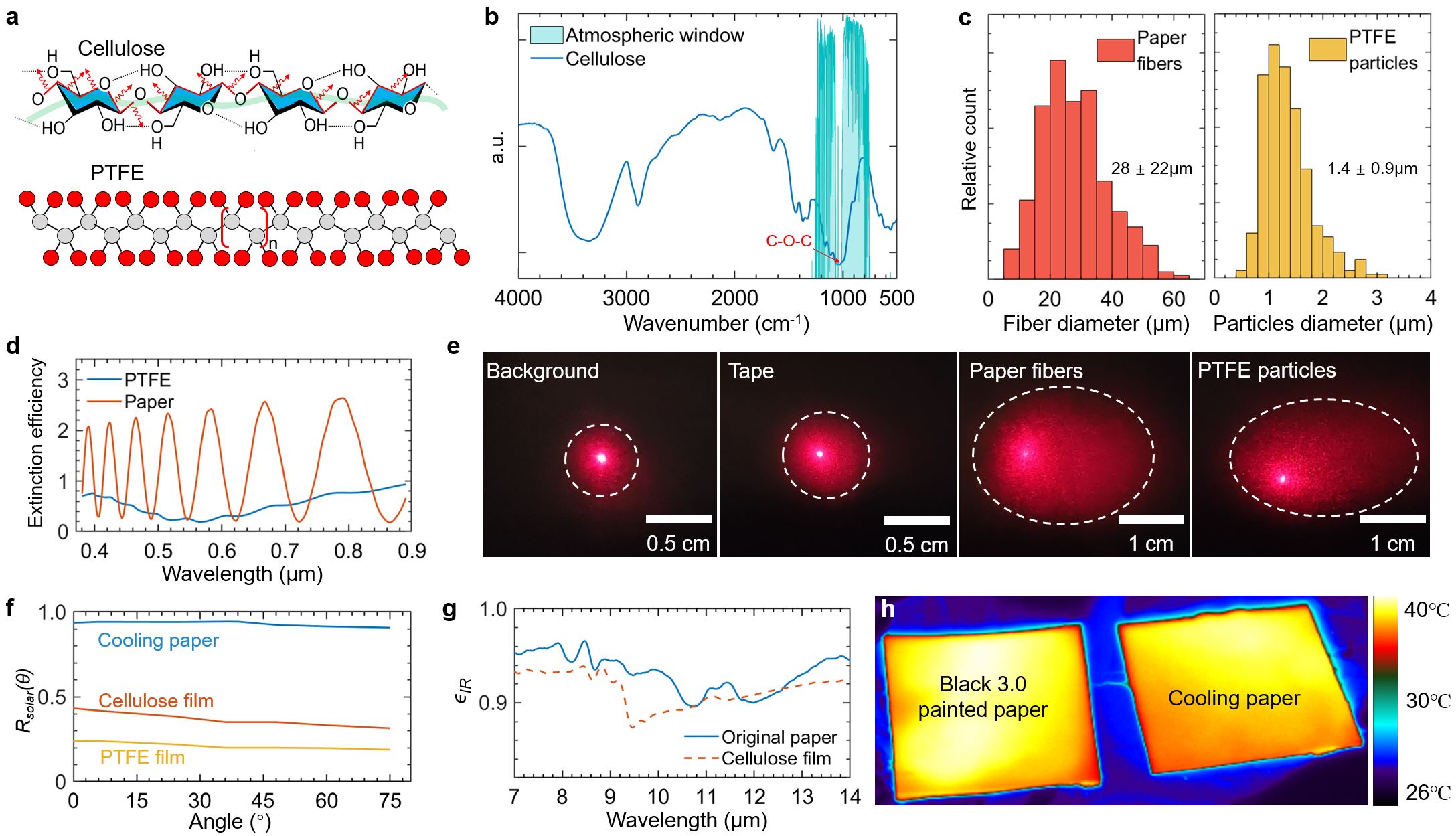}
\caption{\label{fig:physcialmechanism} $\mid$ \textbf{Physical mechanism investigations of advanced optothermal properties of the cooling paper.} \textbf{a}, The schematic showing the molecular structure of cellulose and PTFE (the main components of the cooling paper) and the thermal emissions (red arrows) from the molecular vibrations of C-O-C bands in MIR region. \textbf{b}, FTIR transmittance spectrum of the cellulose presenting C-O-C absorption bands located at the atmospheric transparent window. \textbf{c}, Size distributions of the paper fibers and PTFE particles. \textbf{d}, Simulated scattering cross-section spectra of the paper fibers and PTFE particles that show a high scattering efficiency over solar wavelengths, resulting in a high $R_{solar}$. \textbf{e}, The images showing the scattering effects of a red laser shining at the paper fibers and PTFE particles sticking on a tape film. The clear tape is biaxially oriented polypropylene films, which barely scatters the red laser beam. \textbf{f}, Solar reflectivity ($R_{solar}$) of the cooling paper for different incident angles from 0 to 75$^\circ$ compared with the solid cellulose and PTFE films with the same thickness (700 $\mu$m). \textbf{g}, The infrared emissivity ($\epsilon_{IR}$) of the original paper and the cellulose film from 7 $\mu$m to 14 $\mu$m. \textbf{h}, The infrared images of a Black 3.0 (infrared-ultrablack materials) painted paper and a cooling paper sitting on top of a hot plate. The cooling paper shows high $\epsilon_{IR}$ similar to Black 3.0 painted paper. The deep blue area is the aluminum foil that has little thermal emission in the MIR region.
} 
\end{figure}

Cellulose and PTFE (Fig. \ref{fig:physcialmechanism}a) are two major components of the cooling paper and they own intrinsic optical properties for the PDRC applications. The Fourier transform infrared (FTIR) transmission spectra show that cellulose, lignin and PTFE in the cooling paper exhibit strong emission bands at  800$-$1200 cm$^{-1}$ and 1040$-$cm$^{-1}$, which are assigned to vibrations of C-O-C of cellulose, the vibrations of syringyl rings, guaiacyl rings of lignin and asymmetric and symmetric stretching bands of CF$_2$ of the PTFE (Fig. \ref{fig:physcialmechanism}b, and Supplementary Fig. 8) \cite{abderrahim2015kinetic,smith2011fundamentals,chen2016enhancing}. This result implies that the infrared emission bands of the cooling paper coincidentally fall in the atmospheric transparent window. The complicated organic components (e.g., hemicellulose, lignin, and PTFE) make the paper own wider infrared emission bands and thus the paper exhibit higher $\epsilon_{IR}$ than pure cellulose film. The negligible extinction coefficient of the cellulose and PTFE from 0.38 $\mu$m to 0.9 $\mu$m (covering half of the solar irradiance) makes the cooling paper an ideal reflector of the solar irradiance (Supplementary Fig. 9a,b). The innate structure of the cooling paper with PTFE microparticles and cluttered cellulose fibers presents high $R_{solar}$ and $\epsilon_{IR}$. The diameter of cellulose fibers ranges from 6 $\mu$m to 50 $\mu$m centering at 28 $\mu$m, while the PTFE particles span within 1.4 $\pm$ 0.9 $\mu$m (Fig. \ref{fig:physcialmechanism}c). The cellulose microfibers efficiently backscatter the sunlight, which is further enhanced by adding sub-micron PTFE particles, as shown by the cross-section scattering efficiency spectra simulated using the finite-difference time-domain (FDTD) methods (Fig. \ref{fig:physcialmechanism}d, and Supplementary Fig. 8c). The scattering effects of the cellulose fibers and PTFE particles are visualized using a red laser beam with a wavelength of 650 nm traveling through the cellulose fibers and PTFE particles sticking on a clear tape film (Fig. \ref{fig:physcialmechanism}e). The illuminated area is about 4 times larger than that of the one of the clear tape films indicating strong scattering effects from the paper fibers and particles, similar phenomena are observed for the green and violet laser beams (Supplementary Fig. 10). The disordered cellulose fibers and PTFE particles result in a diffused white surface compared to the solid cellulose and PTFE films (Fig. \ref{fig:physcialmechanism}f and Supplementary Fig. 11) with an AOI from 0 to 75$^\circ$. The cooling paper has a higher infrared emittance than the cellulose film, which is attributed to its components and the random-cluttered cellulose fibers (Fig. \ref{fig:mainspectrum}c,d, Fig. \ref{fig:physcialmechanism}g and Supplementary Fig. 2,3). The high $\epsilon_{IR}$ of the cooling paper is comparable to that of an infrared-ultrablack paint (Black 3.0, $\epsilon_{IR}$ $\approx$ 0.99), which is validated by the infrared images and reflectivity spectra as shown in Fig. \ref{fig:physcialmechanism}h and Supplementary Fig. 12.

\begin{figure}[!ht]
\centering
\includegraphics[width=0.9\textwidth]{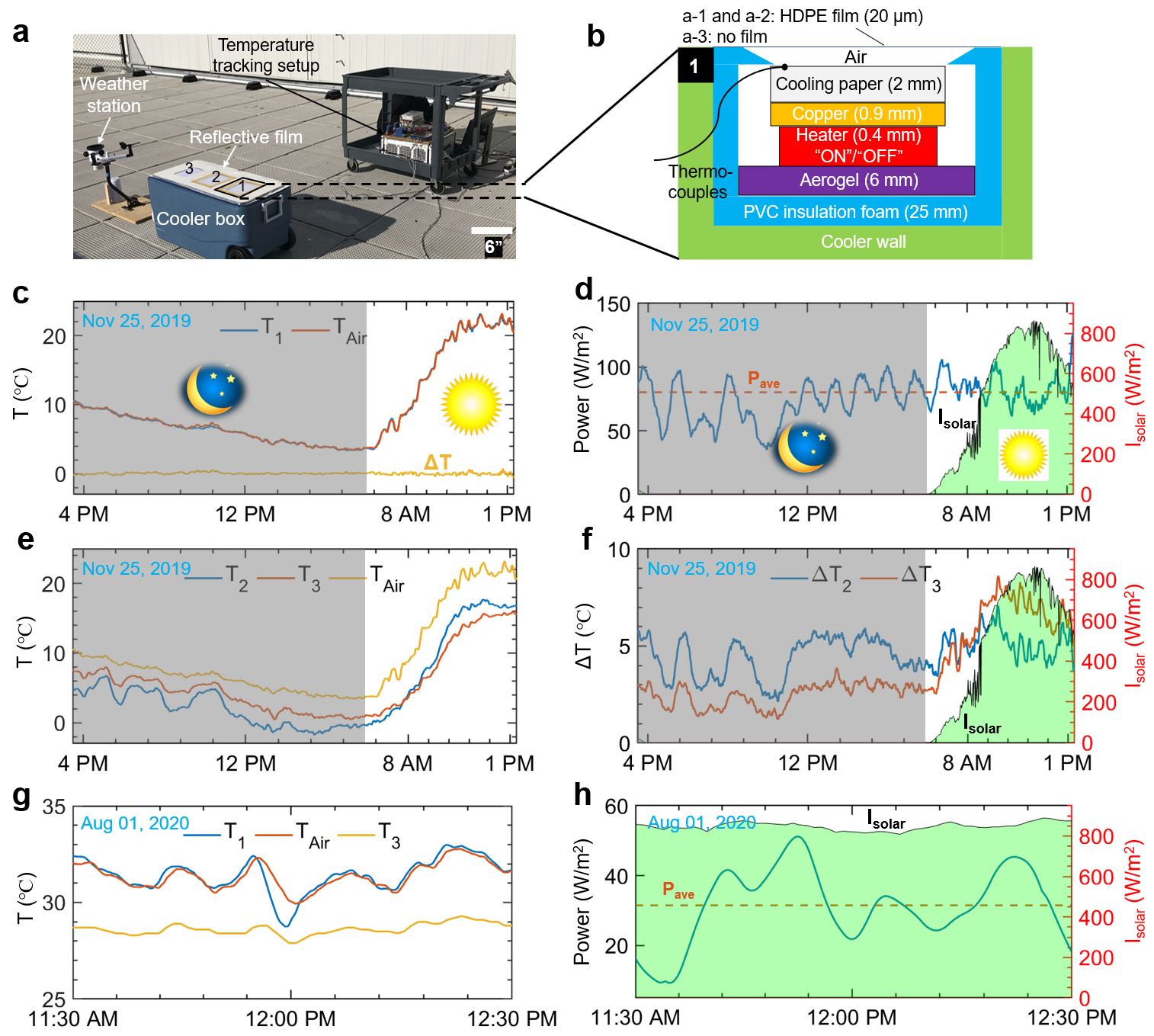}
\caption{\label{fig:outdoorexp} $\mid$ \textbf{Radiative cooling performance of the cooling paper.} \textbf{a}, \textbf{b}, Photograph (\textbf{a}) and schematic (\textbf{b}) of the setup in chamber \textbf{a}-1 for the real-time measurement of sub-ambient cooling performance under sunlight. The test sample is placed in a PVC insulation foam box with bottom-insulated by the aerogel layer to minimize heat leaking. (T$_i$ refers to the surface temperature of sample \textbf{a}-i) \textbf{c}, \textbf{d}, Temperature tracking (\textbf{c}) and cooling power (\textbf{d}) of the cooling paper in one-day cycle. \textbf{e}, Temperature variations of the cooling paper in chambers \textbf{a}-2, \textbf{a}-3 and the ambient. \textbf{f}, Temperature difference between the ambient and the cooling paper in \textbf{a}-2 and \textbf{a}-3, respectively. \textbf{g}, \textbf{h}, Temperature tracking (\textbf{g}) and cooling power (\textbf{h}) of the cooling paper in mid-summer. 
} 
\end{figure}

\section*{Outdoor validations of the self-cooling paper cooling performance}

The sub-ambient cooling performance of the cooling paper during a one-day cycle has been demonstrated on a rooftop at Northeastern University, Boston, MA, USA. A six-layer paper (opaque to the sunlight) is glued layer by layer as the test sample with the top layer sprayed by the PTFE microparticles (Supplementary Fig. 13). Three cooling paper samples (dimensions: 12 cm $\times$ 13 cm $\times$ 2 mm, details of physical properties are listed in Supplementary Table 1) are thermally insulated in three containers insider a modified cooler box. They are used to track and determine the effective radiative cooling power by employing the proportional-integral-derivative (PID) controlled heating compensation system (Fig. \ref{fig:outdoorexp}a-1 and b) and the sub-ambient cooling temperature drops with or without the HDPE windshield, respectively (Fig. \ref{fig:outdoorexp}a-2, a-3 and Supplementary Fig. 14). The top surface of the sample faces to the sky through a high density polyethylene (HDPE) film as a windshield to reduce the air convection effect. The top-surface temperature of the sample is tracked to the ambient temperature by the heat flux from the thin film heater controlled by a PID algorithm. Hence, the heating power of the heater as compensation can be regarded as the radiative cooling power of the sample if an only negligible temperature difference exists between the top surface and the ambient air. The setup in \textbf{a}-2 and \textbf{a}-3 is the same as that in \textbf{a}-1, while the thin-film heater is off. Chambers \textbf{a}-2 and \textbf{a}-3 are for the steady-state temperature recording of the cooling paper with a HDPE windshield (\textbf{a}-2) and without it (\textbf{a}-3). Details of \textbf{a}-2 and \textbf{a}-3 are depicted in Supplementary Fig. 14. The temperature tracking curve shown in Fig. \ref{fig:outdoorexp}c indicates that the top surface temperature of the cooling paper follows the ambient temperature with a bias of $\pm$ 0.2$^\circ$C (Supplementary Fig. 15,16), indicating a reliable measurement accuracy of the radiative cooling performance of the cooling paper. The cooling paper has an averaged radiative cooling power of about 75 W/m$^2$ over a one-day cycle. The average cooling powers of the cooling paper are 82.3 W/m$^2$ and 71.6 W/m$^2$ at the daytime and nighttime, respectively (Fig. \ref{fig:outdoorexp}d). This is attributed to that the nighttime relative humidity (70\% RH, averaged, low atmospheric transmittance) is higher than the daytime one (55.3\%, averaged, high atmospheric transmittance) and the nighttime wind speed (1.9 km/h, averaged, low air convection) is much lower than the daytime one (5.4 km/h, averaged, high air convection) (Supplementary Fig. 17). The $\Delta$T$_2$ (4.4$^\circ$C) is higher than $\Delta$T$_3$ (2.4$^\circ$C) during the night, vice versa during the day ($\Delta$T$_2$ = 5.1$^\circ$C $<$ $\Delta$T$_3$ = 6.1$^\circ$C) (Fig. \ref{fig:outdoorexp}e,f), because the HDPE windshield blocks the heating from the air at night resulting in lower temperatures from the radiative cooling effect. However, the cooling effect from the high-speed wind is partially blocked by the HDPE film in \textbf{a}-2 compared with the one in \textbf{a}-3. The various external factors (solar intensity, non-radiative heat transfer, and atmospheric transmittance) are discussed in Supplementary Fig. 18,19 and Supplementary Notes. Compared with the common building materials such as wood and concrete, the temperature of the cooling paper is 8.1$^\circ$C lower than the one made of basswood (Supplementary Fig. 20,21), attributing to the self-cooling function and the lower thermal conductivity of the cooling paper than that of the basswood and concrete (Supplementary Fig. 22-24). As shown in Fig. \ref{fig:outdoorexp}g,h, the cooling paper in \textbf{a}-3 without a windshield is 2.8$^\circ$C cooler than the ambient, and has an averaged 32 W/m$^2$ cooling power at noontime in the mid-summer (Supplementary Fig. 21).  

\section*{Surface stability of the cooling paper}

\begin{figure}[!ht]
\centering
\includegraphics[width=0.9\textwidth]{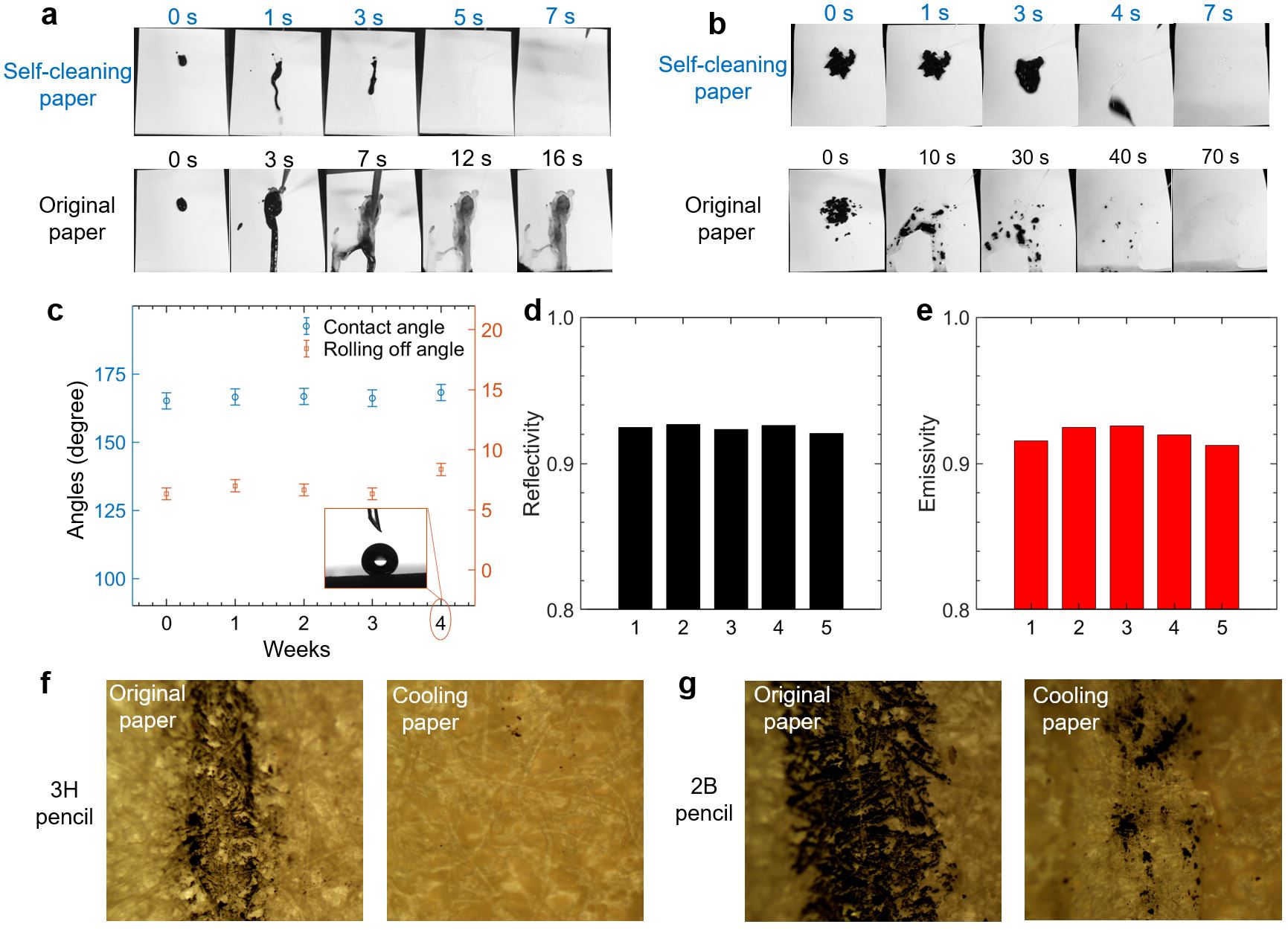}
\caption{\label{fig:selfclean} $\mid$ \textbf{Surface stability of the cooling paper under various simulated environment.} The self-cleaning properties for the cooling paper when stained by black dye (\textbf{a}) and garden soil (\textbf{b}). \textbf{c}, The contact angle and rolling off angle of the cooling paper after outdoor exposure for different time. Inset shows its water contact angle after a four-week outdoor test. \textbf{d}, \textbf{e}, Overall solar reflectance and thermal emissivity of the cooling suffering from various harsh environment for 30 days: 1, pristine cooling paper, 2, -20$^\circ$C, relativity humidity (RH) = 30\%, 3. 60$^\circ$C, RH = 100\%. 4. 80$^\circ$C, RH = 0\%, and 5. UV light (3 mW/cm$^2$). \textbf{f}, \textbf{g}, The micrographs showing the anti-scratch properties of the cooling paper compared to the original one when scratched by the (\textbf{f}) 3H and (\textbf{g}) 2B pencils.
} 
\end{figure}

The cooling paper is mechanically robust because of hydrogen bond interactions and friction forces between fibers (Supplementary Fig. 25) \cite{przybysz2016contribution}. It can be easily cemented to common materials like brick, metal, wood, and concrete for outdoor utilization (Supplementary Fig. 26).  The cooling paper often suffers complicated environmental factors such as dust contamination, UV exposure, temperature and humidity change, mechanical abrasions, raindrop washing, and scratches, any of which may degrade its optical properties and cooling performance in an outdoor environment (Supplementary Fig. 1). This issue is usually neglected in past investigations. We examine the long-term stability of the cooling paper under a simulated environment (Fig. \ref{fig:selfclean}a-e and Supplementary Fig. 27-41). Black dye and garden soil are employed to contaminate both original paper and cooling paper. The original paper is easily dyed by black dye and the soil on its surface is not easily washed away, resulting in reduced light reflectivity. In contrast, the cooling paper remains clean without any dying and the residual soil because of its excellent self-cleaning ability arising from the superhydrophobic PTFE coating (Fig. \ref{fig:selfclean}a,b). The self-cleaning durability of the cooling paper is demonstrated by the four-week outdoor exposure with little changes observed in the contact angle and rolling off angle (Fig. \ref{fig:selfclean}c, and Supplementary Fig. 28-30). The 15-minute water immersion test shows the good water-proof properties for outdoor utilization (Supplementary Fig. 31). Additionally, different harsh environmental tests, such as UV exposure, rain-drop impact, low and high-temperature exposure, high humidity exposure, and mechanical abrasions do not lead to any change of optical performance and self-cleaning property (Fig. \ref{fig:selfclean}d,e, and Supplementary Fig. 32-38). This indicates good surface stability of the cooling paper, which is a benefit to keeping its good cooling performance for outdoor applications. The PTFE layer of the cooling paper can also reduce the friction of the cooling paper surface, avoiding scratches from hard objects (Fig. \ref{fig:selfclean}g,h, and Supplementary Fig. 39,40). After cyclic peel-off tests, the reflectivity and emissivity of the cooling paper have no significant change (both are within 3\%), demonstrating its robust cooling performance (Supplementary Fig. 41,42). It is also experimentally demonstrated that the recycled cooling paper still yields a self-cooling performance by spectroscopic analysis, elongating its lifespan (Supplementary Fig. 43). 

\begin{figure}[!ht]
\centering
\includegraphics[width=1\textwidth]{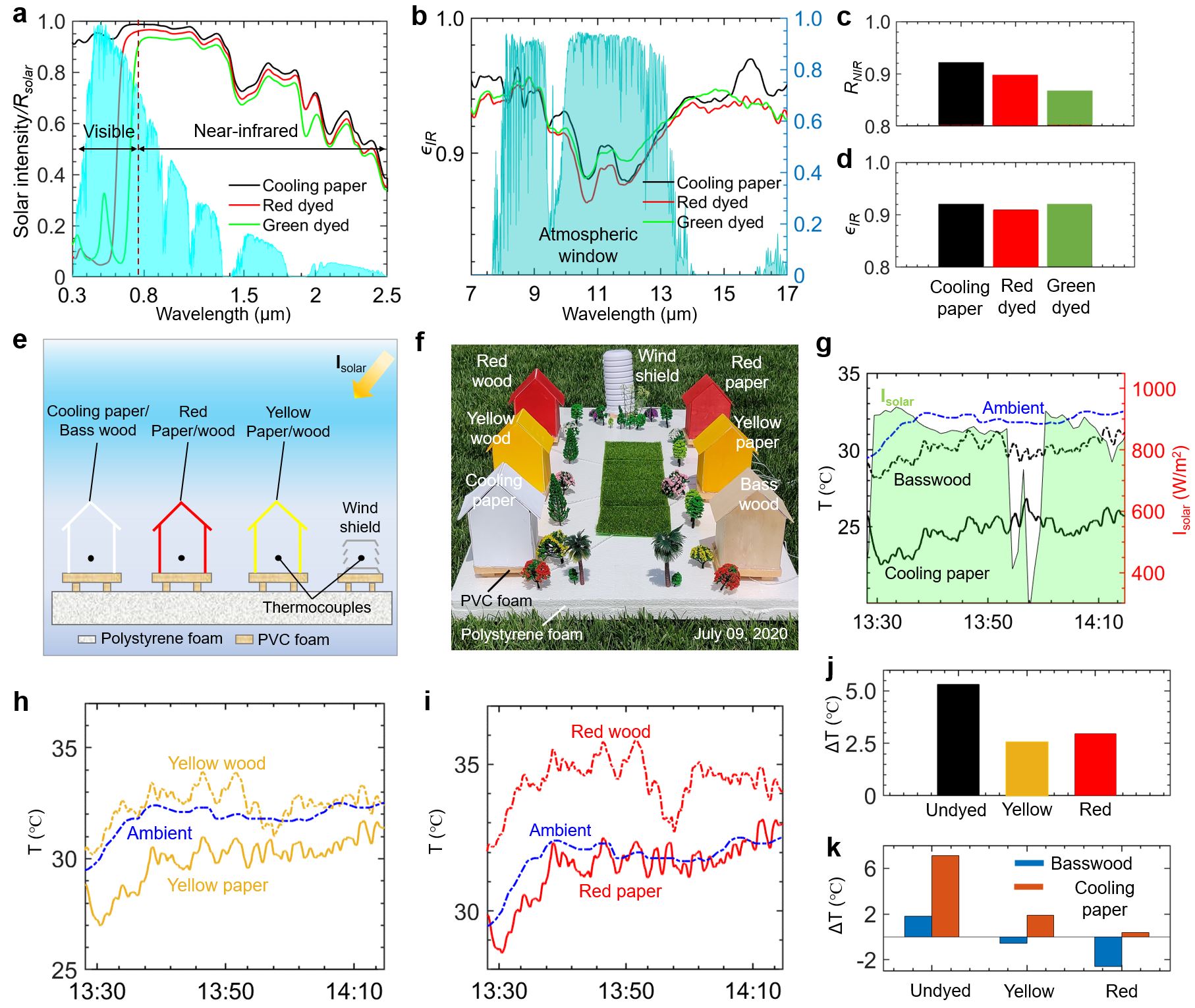}
\caption{\label{fig:dyedpaper} $\mid$ \textbf{The cooling performance of the dyed paper.} \textbf{a}, \textbf{b}, $R_{solar}$ (\textbf{a}) and $\epsilon_{IR}$ (\textbf{b}) of the red and green dyed papers compared to the cooling paper. \textbf{c}, \textbf{d}, The averaged near-infrared (NIR) solar reflectivity (\textbf{c}) from 0.75 $\mu$m to 2.5 $\mu$m and $\epsilon_{IR}$ (\textbf{d}) from 8.0 $\mu$m to 13 $\mu$m of the cooling paper, red and green dyed papers. \textbf{e}, \textbf{f}, Schematic (\textbf{e}) and photograph (\textbf{f}) of the setup for the temperature recording under sunlight. \textbf{g}, \textbf{h}, \textbf{i}, Temperature variations (left axis) and solar intensity (right axis) in the outdoor test of the basswood and cooling paper (\textbf{g}), yellow wood and paper (\textbf{h}), and red wood and paper (\textbf{i}) for 45-minute test at noontime in the mid-summer. \textbf{j}, The average temperature difference of cooling paper and the basswood during the test. The temperature of the cooling paper is lower than the basswood both undyed and yellow/red-dyed. \textbf{k}, The averaged temperature difference between the ambient and the cooling paper and basswood (undyed, yellow and red-dyed). Positive values show the chamber temperature made of the cooling paper or basswood is lower the ambient air, vice versa. 
} 
\end{figure}

\section*{Colorful paper with efficient cooling performance and aesthetics}

The compatibility with self-cooling functionality and aesthetics requires that a surface reflects a certain wavelength band in the visible region to display colors, meanwhile maximizes its $R_{solar}$ in near-infrared (NIR) wavelengths (0.7$-$2.5 $\mu$m), where 51\% of the solar energy lies to minimize the solar heating, and owns a high $\epsilon_{IR}$ in the atmospheric window (8 $-$ 13 $\mu$m) to radiate heat efficiently to the cold outer space. Red- and green-dyed papers can efficiently reflect the NIR sunlight with a high $R_{NIR}$ of 0.89 and 0.87, respectively, to prevent from being heated up by the NIR solar wavelengths (Fig. \ref{fig:dyedpaper}a-d). To further validate the efficient cooling performance of the white and colorful paper under the hot mid-summer weather, mini houses made of the white, yellow, and red-dyed cooling papers are fabricated and their "room" temperatures are recorded during the noontime (13:30 to 14:15) on July 9th, 2020. The mini houses made of basswood with different colors are selected as control groups (Fig. \ref{fig:dyedpaper}e,f, and Supplementary Fig. 44,45). We find that the white house made of the cooling paper is 7.1$^\circ$C cooler than the ambient air and even 5.3$^\circ$C lower than that of the basswood-made house (Fig. \ref{fig:dyedpaper}g). Similar trends of the cooling performance are observed in the yellow and red mini houses made of dyed cooling papers and corresponding basswood (Fig. \ref{fig:dyedpaper}h-k). The dyed cooling paper will provide a promising alternative for the colored radiative cooling utilization, greatly expanding its scope of applications.

\section*{Conclusions}
We demonstrate a self-cleaning and self-cooling paper made of cellulose fiber-based paper and PTFE particles that can be scalable-fabricated and easily recycled for the PDRC applications. The cooling paper is super ``white" in the solar wavelengths, resulting from the backscatter of sunlight by the randomized structure of the cellulose microfibers and PTFE microparticles, while it is infrared ``black" in the atmospheric window because of the molecular vibrations of chemical bonds in cellulose and PTFE. The super ``white" and ``black" features at different wavelengths ensure that the cooling paper has a net radiative cooling loss to the cold outer space. Its performance is better than or on par with other radiative cooling performances as listed in Supplementary Table 2. The outdoor experiment demonstrates the excellent cooling effect in both white and colored forms of the cooling paper. Moreover, the cooling papers possess a self-cleaning and robust surface, which is a benefit to keeping its good cooling performance for the outdoor environment without labor-intensive maintenance. This multifunctional cooling paper provides a promising pathway for practical application in energy-saving and sustainable buildings. 

% \newpage
\section*{Methods}
% \noindent
\subsection*{Materials}
The 100 lb paper was provided by INTERNATIONAL PAPER. The spray adhesive was from GORILLA. The PTFE film with a thickness of 20 mils was purchased from CS Hyde company. Black dye was from AmeriColor corporation. Red and yellow dyes were from Rite LLC. All reagents were used as received without future purification. 

\subsection*{Fabrication of the cooling paper}
\textbf{\textit{Fabrication of the white cooling paper.}} Three grams of polytetrafluoroethylene (PTFE) microparticles (1.4 $\mu$m average size in diameter, Sigma-Aldrich) were added into 30g ethanol (95\% denatured, Innovating Science) and homogenized by a high-speed emulsifier at 22,000 rpm for 3 minutes. The resultant suspension was sprayed on to a 12 cm $\times$ 13 cm paper by airbrush equipment (nozzle diameter 0.8 mm) at an air pressure of around 70 psi air pressure with a distance of about 10 cm. The mass density of PTFE coating on the paper is around 2 mg/cm$^2$. The sprayed paper was dried at 100$^\circ$C for 30 minutes.

\noindent
\textbf{\textit{Fabrication of the colored cooling paper.}}
5g dye (Rite LLC) was dissolved into 10g DI water and was subsequently sprayed onto a 12 cm $\times$ 13 cm paper by the same airbrush using the same parameters with the fabrication of the cooling paper. The paper can also be dyed by dipping it into the dye suspension. The sprayed or dip-coated paper was then dried at 60$^\circ$C for 30 minutes.  

\noindent
\textbf{\textit{Fabrication of Solid cellulose film.}}
Two grams of dried cellulose powders were placed into a dry pellet pressing die with an inner diameter of 13 mm and was pressed at a pressure of 60 MPa for 45s to form a uniform solid film.

\noindent
\textbf{\textit{Recycling process of the cooling paper.}}
The cooling paper was cut into small pieces and was immersed into water for 1 hour. After that, the swelled paper was smashed into paper pulp using the high-speed blender. The paper pulp was thermally-pressed into a paper sheet by the hydraulic press machine under 20 MPa and 200$^\circ$C until the paper is dry. The regenerated paper can be reused for fabricating the cooling paper.

\subsection*{Material characterization}
\textbf{\textit{Morphology characterization.}}
The surface morphologies of samples were examined by scanning electron microscopy (SEM, S5200, Hitachi Company) under an acceleration voltage of 10 kV. The morphology of the pencil-scratched samples was observed by a Trinocular Metallurgical Microscope (ME300TZA-3M) with a 50X lens.

\noindent
\textbf{\textit{Surface wetting characterization.}}The water contact angle (CA) of the sample was measured by a contact angle goniometer (DSA-25, Kruss, Germany) with 7 $\mu$l DI water droplet at room temperature according to the sessile droplet method. The reported static contact angles were calculated by averaging the measured values from both the left and right sides of the droplet. Six data points were collected at six different positions on the sample surface and were used for calculating the final average values. 

\noindent
\textbf{\textit{Paper composition characterization.}}
1g ($\pm$ 0.0001g) cooling paper sample was dried at 100$^\circ$C for 30 minutes and was weighed to calculate the moisture content. The PTFE content was identified by measuring the mass change of the paper before and after spraying PTFE. 400ml ethanol (95\%) was mixed with 100ml HNO$_3$ for etching other components of the paper to obtain the cellulose. 1g ($\pm$ 0.0001g) cooling paper sample was mixed with 25g ethanol-HNO$_3$ solution in a 250ml conical flask with a condenser and was subsequently refluxed for 1 hour at 100$^\circ$C. These procedures were repeated 2 to 3 times until the paper fibers were white. The above fibers were washed using the ethanol until the pH of the filtrate is near to 7. Finally, these fibers were dried at 105$^\circ$C ($\pm$ 2$^\circ$C) and were weighted for calculating of the cellulose content.

\noindent
\textbf{\textit{Optical characterization.}}
The reflectivity spectra (0.3 $\mu$m $\sim$ 2.5 $\mu$m) were measured by the Jasco V770 spectrophotometer at an incident angle of 6$^\circ$ with the ISN-923 60 mm BaSO$_4$ based integrating sphere equipped with PMT and PbS detectors. The reflectivity spectra are normalized by a PTFE based reflectance standard. The reflectivity spectra (2.5 $\mu$m $\sim$ 20 $\mu$m) were measured by Jasco FTIR 6600 at an incident angle of 12$^\circ$ with the PIKE upward gold integrating sphere equipped with wide-band MCT detector. Angular-dependent reflectivity spectra were measured by using wedges with different angles at the sample port of these two integrating spheres. 

\noindent
\textbf{\textit{Refractive index characterization.}}
The complex refractive index of the sample was measured by ellipsometry (J.A.Woollam, M-2000DI) with an incident angle of 60$-$70$^\circ$ and spectral range of 0.38$-$0.9 $\mu$m.

\noindent
\textbf{\textit{Size distribution characterization.}}
The size distribution of the cellulose fibers and PTFE particles was determined by the ImageJ software according to the SEM images, in which 200 points were randomly selected for each image.

\noindent
\textbf{\textit{FDTD simulation.}}
FDTD simulation was executed using the Lumerical FDTD Solution 2018a. Two-dimensional models were employed and a total-field scattered-field source coupled with two scattering cross-sections of the cellulose fibers and PTFE particles were used to calculate the scattering efficiency.

\noindent
\textbf{\textit{Scattering effects measurement.}}
One layer of paper fibers was peeled off from the original paper using the Tape King packaging tape. 0.2g PTFE particles were uniformly smeared on the adhesive side of the packing tape (CS Hyde company). These cellulose fibers and PTFE particles were placed at the aperture position, which is between the laser pens (red, green, and purple) and the blackboard. The area of scattered laser spots was used to indicating the scattering effect of paper fibers and PTFE particles.

\noindent
\textbf{\textit{Infrared image measurement.}}
Infrared images were taken using FLIT A655C thermal camera with 25$^\circ$ lens at a resolution of 640 $\times$ 480.

\noindent
\textbf{\textit{Cooling power and temperature tracking.}}
The temperature of the cooling paper sample was measured using the K-type thermocouples connected to the National Instruments (NI) PXI-6289 multifunction I/O module. The ``ON" and ``OFF" of the heater was controlled by the NI PXI-2586 relay module driven by a home-built LabVIEW program using the PID control algorithm to maintain the temperature of the cooling paper tracking the ambient air.

\noindent
\textbf{\textit{Thermal conductivity measurement.}}
The thermal conductivity of one layer paper was characterized by TPS 2500s thin-film module, while the thermal conductivities of the basswood and concrete were measured by the isotropic standard module of TPS 2500s.

\noindent
\textbf{\textit{Mechanical strength measurement.}}
The mechanical strength of the paper sheet (2 cm $\times$ 6 cm $\times$ 0.35 mm) and a basswood slab (2 cm $\times$ 6 cm $\times$ 1.5 mm) were measured by the Mark-10 ESM tensile tester at room temperature.

\noindent
\textbf{\textit{Abrasion robustness tests.}}
A sandpaper abrasion test is carried out using 400 grit SiC sandpaper as an abrasion surface. The sample was fixed to a glass slide with a loaded weight of 50 g and placed face-down onto sandpaper, and then moved 10 cm along the ruler. Subsequently, the sample was rotated by 90$^{\circ}$ and again moved 10 cm along the ruler \cite{li2019robust}. The two aforementioned steps are defined as one abrasion cycle. This process guarantees that the cooling paper surface is abraded longitudinally and transversely in each cycle.

\noindent
\textbf{\textit{Anti-scratch tests.}}
A set of pencils with different hardness (10B-9B-8B-7B-6B-5B-4B-3B-2B-B-F-HB-H-2H-3H-4H-5H-6H-7H-8H-9H) were selected to make a scratch on the cooling paper surface. The pencil was held at 45$^\circ$ to the cooling paper and pushed for a scratch length of 10.0 mm. The micrographs of the paper after scratch tests were used to analyze its anti-scratch property.

\noindent
\textbf{\textit{Rain-drop tests.}}
The simulated rain-drop tests were conducted using a syringe pump at a flow rate of 3ml/h from the 17 gauge syringe needle for 7 days. The needle head is about 35 cm over the cooling paper. The oblique angle of the cooling paper is fixed at 10$^\circ$.

\noindent
\textbf{\textit{Peel-off tests.}}
An XFasten tape (6 cm $\times$ 2 cm) was pressed on to the cooling paper and was peeled off at a speed of about 1 cm/s.

\noindent
\textbf{\textit{Harsh environment exposure.}}
To test the long-term durability of the cooling paper under different harsh environment exposure, the cooling papers were placed in a freezer ($-$20$^\circ$C), an incubator (60$^\circ$C, RH $\approx$ 100\%), a convection oven (80$^\circ$C, RH $\approx$ 0\%), and a UV lamp (UV density, 3 mW/cm$^2$) for 30 days, respectively.

% \newpage
\section*{Data availability}
The datasets analyzed during the current study are available
within the paper. All other data related to this work are available from the corresponding author upon reasonable request.

% \newpage
% \bibliographystyle{ieeetr}
\bibliographystyle{naturemag}
\bibliography{Yanpei}

% \newpage
\section*{Acknowledgments}
This project is supported by the National Science Foundation through grant number CBET-1941743 and the National Natural Science Foundation of China through grant number 51973245.

\section*{Author contributions}
Y.T., S.H., and X.L. developed the concept of scalable self-cleaning paper for radiative cooling. Y.T., X.L., and F.C. conducted self-cooling related experiments and optothermal and mechanical tests. S.H. and Y. L. conducted self-cleaning related experiments. Y.T. wrote the manuscript with help from all other authors. C.T., and Y.Z. contributed to the development of ideas and approaches. All authors provided critical feedback and contribute to the writing of the manuscript. C.T. and Y.Z. supervised this project.

\section*{Competing interests}
The authors declare no conflict of interest.

\section*{Correspondence}
Correspondence and requests for materials should be addressed to C.T. and Y.Z.

\end{document}